\documentclass[baaa]{baaa}

\usepackage[pdftex]{hyperref}
\usepackage{subfigure}
\usepackage{natbib}
\usepackage{helvet,soul}
\usepackage[font=small]{caption}

\contriblanguage{1}

\contribtype{1}

\thematicarea{1}

\title{Dust emission from the first massive galaxies}

\titlerunning{Dust emission from the first massive galaxies}

\author{M.E. De Rossi\inst{1,2}, G.H. Rieke\inst{3}, I. Shivaei\inst{3}, 
V. Bromm\inst{4} \& J. Lyu\inst{3}}
\authorrunning{De Rossi et al.}

\contact{mariaemilia.dr@gmail.com, ghrieke@gmail.com}

\institute{Universidad de Buenos Aires, Facultad de Ciencias Exactas y Naturales y Ciclo B\'asico Com\'un. Buenos Aires, Argentina
\and CONICET-Universidad de Buenos Aires, Instituto de Astronom\'{\i}a y F\'{\i}sica del Espacio (IAFE). Buenos Aires, Argentina
\and Steward Observatory, Department of Astronomy, University of Arizona, 933 North Cherry Avenue, Tucson, AZ 85721
\and Department of Astronomy, University of Texas at Austin, 2511 Speedway, Austin, TX 78712, USA
}

\resumen{
	Comparamos las distribuciones espectrales de energía (SEDs, por sus siglas en inglés) de galaxias observadas en 
	el infrarojo
	lejano (FIR, por sus siglas en inglés) a $z\sim6$ y las predicciones teóricas para las primeras galaxias 
	masivas de población II (Pop II, por sus siglas en inglés).  Las SEDs en el FIR a $z\gtrsim5$ son
	anchas y están desplazadas hacia longitudes de onda más azules comparadas con galaxias a $z\sim3$.
	Mediante la implementación de un modelo analítico para la emisión por polvo de galaxias Pop II, pudimos reproducir
	el comportamiento observado como consecuencia de las altas densidades de energía y la composición de polvo
	rica en silicio de las galaxias modeladas a alto $z$.  Como se notó en un trabajo previo, la falta de naturaleza de cuerpo
	negro de las SEDs de galaxias a $z \sim 6$ debería tenerse en cuenta cuando se interpretan las mediciones
	de las luminosidades en el FIR para evitar subestimar las tasas de formación estelar.
	}

\abstract{
	We compare observed far-infrared (FIR) galaxy spectral energy distributions (SEDs) at
	$z\sim6$ and theoretical predictions for first massive population II (Pop II) galaxies. Observed FIR SEDs at
	$z\gtrsim5$ are broad and shifted to bluer wavelengths when compared to galaxies at $z\sim3$.
	By implementing an analytical model for dust emission from Pop II massive galaxies, 
	we were able to reproduce the observed behaviour as a consequence of the high energy densities and 
	silicate-rich dust composition of high-$z$ model galaxies. As noted in a previous work, the 
	non-blackbody nature of galaxy SEDs at $z \sim 6$ should be taken into account when interpreting
	measurements of FIR luminosities  to avoid underestimating star formation rates.
}


\begin{document}

\maketitle

\section{Introduction}
\label{sec:Introduction}
The behaviour of the far-infrared (FIR) spectral energy distributions
(SEDs) of star-forming (SF) galaxies at high redshifts ($z\gtrsim3$) is 
under debate in the literature. Given the relatively little information available, 
these SEDs have been characterized in simplified ways.
By using indirect arguments, \citet{faisst2017} concluded that galaxies
at very high $z$ show generally "hotter" SEDs than galaxies in the 
Local Universe ($0 \le z \lesssim 0.1$).
According to \citet{lyu2016}, after subtracting the AGN contribution, the host FIR SEDs of $z\sim$5--6 quasars are better described
by the relatively warm SED of the metal-poor starbursting
galaxy Haro 11 ($z \approx 0.02$), instead of the normal metal-rich SF templates used at $z\lesssim2-3$ or modified blackbodies.

\citet{derossi2018} modelled the FIR SEDs corresponding to massive Population (Pop) II galaxies
(formed after the Pop III stage) during the first phase of significant star formation,
finding that these SEDs are significantly shifted to bluer ("warmer") wavelengths 
compared to the SEDs of local galaxies. In addition, adopting a silicate-rich interstellar dust composition with a small percentage of carbon dust, the theoretical SEDs have very similar behavior to 
that of Haro 11. Given its moderately-low metallicity and very young stellar population, 
Haro 11 probably describes the relevant properties in young massive Pop II galaxies at high $z$.
In particular, \citet{derossi2018} reported a progression with $z$ in the SEDs of observed galaxies,
from those resembling local ones at $2\lesssim z<4$ to a more similar behaviour to Haro 11 at
$5\lesssim z \lesssim 7$.  Such variations should be taken into account when estimating
total infrared luminosities at $z\gtrsim5$ from measurements close to $\lambda \sim 1$ mm.

In this article, we extend the work by \citet{derossi2018} performing a detailed
comparison between the theoretical SED, the observed data and different templates SEDs
at the extreme redshift end ($z=5-7$).

\begin{figure*}[!h]
  \centering
  \includegraphics[width=0.7\textwidth]{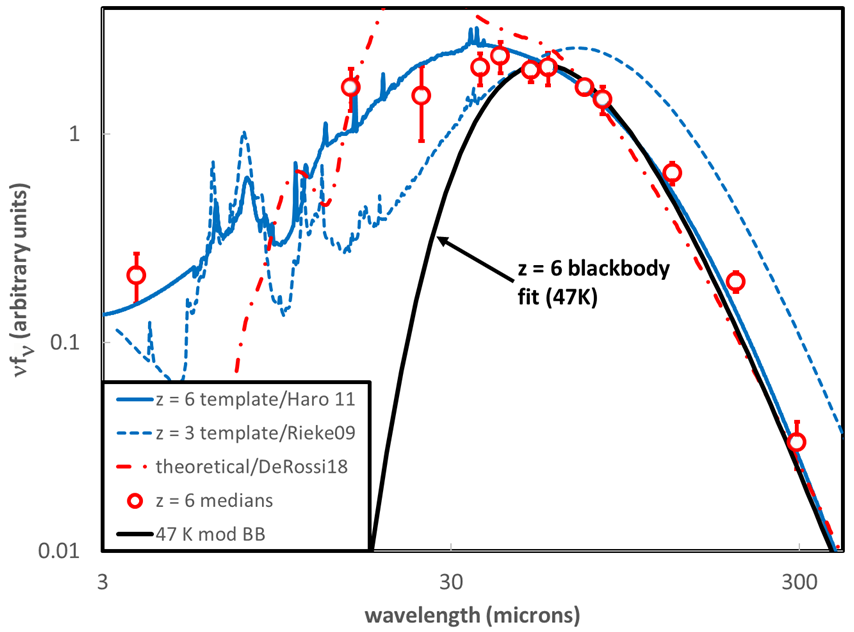}
  \caption{
Comparison between theoretical and observed FIR SEDs at high $z$.  Different curves
and symbols depict medians of the measurements at $z\sim6$ shifted to the rest-frame
(open red circles, 1$\sigma$ error bars), the Haro 11 template (solid blue line), the template that fits best at 
$z = 2 - 4$ (dashed blue line),
	a standard modified blackbody fit for $z = 6$ ($T=47$~K, $\beta = 1.6$, solid black line), 
	and the theoretical model preferred by \citet{derossi2018} (UM-D-20 model for silicate
	dust with 10\% carbon, dot-dashed red line).
  }
  \label{fig:fig1}
\end{figure*}

\section{Dust model}
We adapt the model of \citet{derossi2017} to conditions in
very luminous galaxies 
being detected in the FIR at $z\sim6$ (see \citealt{derossi2018}, for more details).
A model galaxy consists of a dark matter halo hosting a central cluster of Pop~II stars, surrounded by a mixed
phase of gas and dust.
We assume a dust-to-metal mass
ratio $D/M = 0.02$, a gas metallicity of $Z_{\rm g} =0.33~Z_{\odot}$ and a star
formation efficiency of $\eta =0.01$.

We considered the different silicon-based dust models
\citep{cherchneff2010} implemented by \citet{derossi2017}, 
but with 10\%  of the total luminosity contributed by
carbon. We adopted the ‘standard’ grain-size distribution used
in \citet{ji2014}. A dust temperature ($T_{\rm d}$) was determined assuming
thermal equilibrium and the dust emissivity was estimated by applying
the Kirchhoff’s law for the estimated $T_{\rm d}$ profile.

\section{Results}
Fig.~\ref{fig:fig1} compares the theoretical 
model preferred by \citet{derossi2018} (UM-D-20 with 10\% carbon, dot-dashed red line) with
the medians of the measurements at $z\sim6$ shifted to the rest-frame (open red circles, see \citealt{derossi2018} for details 
about individual and stacked observational data). Other curves in Fig.~\ref{fig:fig1} depict
the Haro 11 template (solid blue line), the template that fits best luminous galaxies at $z = 2 - 4$ \citep{rieke2009}
(dashed blue line), and
a standard modified blackbody fit for $z = 6$ ($T=47$~K, $\beta = 1.6$), which is commonly
used to interpret the ALMA $1$~mm measurements. 
The modified blackbody has been fitted to SEDs within the wavelength range accessible to ALMA 
at $z=6$.  To show relative luminosity in logarithmic wavelength bins,
the SEDs are represented by $\nu f_{\nu}$ units with an arbitrary normalization.

We see that observed galaxies at $z\sim6$ have broad SEDs, which are bluer that
those typical at lower $z$. It is clear that observed trends are well reproduced by the continuum predicted by the theoretical model of
\citet{derossi2018}. Departures from the model occur around 20 $\mu$m due to silicate emission and at wavelengths $\le 8 \mu$m where hot dust, probably in close proximity to the hot stars, contributes to the observed fluxes. The model behavior is due to 
higher dust temperatures 
reached at higher $z$, caused mainly by the high energy densities inside luminous galaxies. 
In addition,  
their dominant silicate dust has high radiative efficiency at $\lambda = 8-60~\mu$m, augmenting the effects of the high energy density.
The high energy densities inside the first massive galaxies results from the extremely high luminosities of their
young stellar component ($\sim 10^{13}~{\rm L}_{\odot}$) combined
with their compact sizes \citep{spilker2016}.

\begin{figure}[!h]
  \centering
  \includegraphics[width=0.4\textwidth]{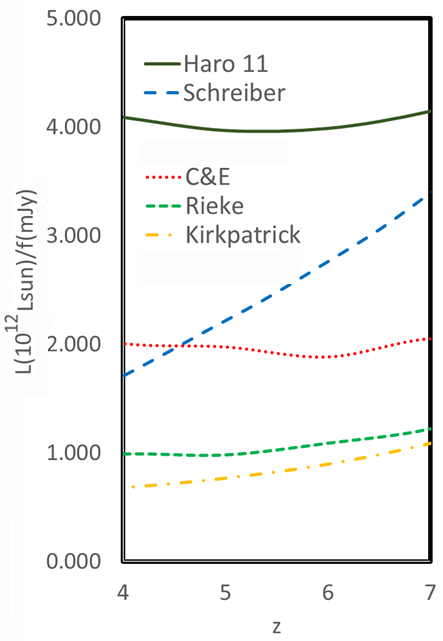}
  \caption{Conversion factors from ALMA oberved Band 7 flux
densities in mJy to L(TIR) in units of $10^{12}$ L$_\odot$, for various SED
templates: \citet{schreiber2018, rieke2009, chary2001, kirkpatrick2015} and also with the Haro 11 template. The \citet{rieke2009} template is applied as in \citet{rujo2013}.
}
  \label{fig:fig2}
\end{figure}

Fig.~\ref{fig:fig1} also shows that the SED template based on the local galaxy Haro 11 
constitutes a good representation of the $z\sim 6$ data and is also well described by the theoretical model.
As discussed in \citet{derossi2018}, this is expected as the interstellar medium conditions in Haro 11 are in
reasonable agreement with those corresponding to luminous infrared galaxies at $z\sim 6$. Thus, 
\citet{derossi2018} proposed the Haro 11 SED as a proxy template to describe the trends of 
FIR SEDs at high $z$.
On the other hand, a modified blackbody fit (solid black line) provides a good representation of the data  
within the range of rest wavelengths accessible with ALMA at submm- and mm-wave observations ($\lambda > 40~\mu$m),  
but leads to an underestimation of total infrared luminosities – and star formation rates –
by a factor of 2. Other commonly used templates also result in substantial underestimates of the infrared luminosities based just on ALMA measurements near 1 mm compared with the estimate using the Haro 11 template, as shown in Figure~\ref{fig:fig2}. Here we plot the ratio of deduced FIR luminosity to the Band 7 flux density (in mJy) as  a function of redshift, as it is predicted by a number of commonly used templates.  

\section{Conclusions}
In this work, we compared observed FIR galaxy SEDs at $z\sim6$
and theoretical predictions for the first massive Pop II galaxies.
Our analysis is based on the dust model developed by \citet{derossi2017}
but adapted to conditions reflecting the properties of luminous
FIR galaxies at $z=5-7$ \citep{derossi2018}.

The observed SEDs at $z\sim6$ are broad and shifted to bluer wavelengths
in comparison with SEDs associated with galaxies at lower $z$.  The theoretical dust model is able
to reproduce the observed trends as a consequence of the high energy densities
of model massive Pop II galaxies and their dominant silicate dust composition.
As discussed in \citet{derossi2018}, the SED corresponding to the local galaxy
Haro 11 constitutes a convenient proxy template to study galaxies at $z\gtrsim 5$.

The characterization of FIR galaxy SEDs at high $z$ is crucial for 
the exploration of the properties of the interstellar medium of high $z$ galaxies.
A straightforward application is the estimate of total FIR luminosities,
which, once combined with the initial mass function, provide an estimate
of the star formation rates in luminous dust-shrouded
galaxies.
Single-band ALMA measurements at  $\sim 1$~mm are being used to infer IR luminosities
and star formation rates at $z \approx 6$. However, the interpretation of such
data depends on the adopted SED template as shown in Figure~\ref{fig:fig2} \citep[see also][]{derossi2018}. 
The non-blackbody nature for galaxy SEDs at $z\gtrsim5$, predicted by our theoretical model and supported by the available measurements, needs to be considered to avoid an underestimation of total infrared luminosities and, hence,
the star formation rates at these high redshifts.

\begin{acknowledgement}
MEDR thanks the Asociación Argentina de Astronomía for providing
with partial financial support for attending its 61st annual meeting.
The work of GHR, IS, and JL was partially supported by NASA Grant NNX13AD82G, and
that of IS was also partially supported by a Hubble Fellowship.
MEDR is grateful to PICT-2015-3125 of ANPCyT (Argentina) and also to
Mar\'{\i}a Sanz and Guadalupe Lucia for their help and support.
VB acknowledges support from NSF grant AST-1413501.
We thank Alexander Ji for providing tabulated dust opacities for the different
dust models used here.
This work makes use of the Yggdrasil code \citep{zackrisson2011}, which adopts
Starburst99 SSP models, based on Padova-AGB tracks \citep{leitherer1999, vazquez2005}
for Population~II stars.
\end{acknowledgement}


\bibliographystyle{baaa}
\small
\bibliography{bibliografia}
 
\end{document}